\documentclass[preprint,12pt]{elsarticle}

\usepackage{amssymb}
\usepackage{lipsum}
\usepackage{amsmath}
\usepackage{xcolor}
\usepackage[colorlinks]{hyperref}
\begin{document}

\begin{frontmatter}



\title{Nonlinearity Mediated Miscibility Dynamics of Mass-imbalanced Binary Bose-Einstein Condensate for Circular Atomtronics}

 \author[label1]{Sriganapathy Raghav}
 \affiliation[label1]{organization={Department of Physics, Indian Institute of Technology},
            city={Patna},
            postcode={801106},
            state={Bihar},
            country={India}}
  
  \author[label1,label2]{Suranjana Ghosh}
  \affiliation[label2]{organization={Department of Physics, Indian Institute of Science Education and Research},
  	city={Kolkata},
  	postcode={741246},
  	state={West Bengal},
  	country={India}}

\author[label1]{Barun Halder}
\author[label1]{Utpal Roy}
\ead{uroy@iitp.ac.in}

\begin{abstract}
We explore the nonlinearity-induced and fractional revivals-driven miscibility dynamics of quasi-2D mass-imbalanced binary Bose-Einstein condensates, confined in a ring-shaped waveguide. During their time-evolution, the two condensate species generally remain miscible, as observed in the spatial density distributions and the autocorrelation functions. Although, the investigation is carried out for a wide range of mass-imbalance, initial demonstration is focussed on insignificant mass-imbalance of the two $Rb$-isotopes with suitable experimental parameters. The characteristic time scales are influenced by the trap parameters and the strengths of nonlinearities. The study also reveals the conditions under which the condensates become spatially distinguishable with clear signatures in their autocorrelation functions. A separability function further identifies favorable parameters and the fractional revival instances for greater separability. We report precise range of the ring-radius and the interaction strength for experimental realization. Additionally, the average separability variation reflects the result across a variety of condensate species.
\end{abstract}




\end{frontmatter}




\section{Introduction}

Binary mixture of Bose-Einstein condensates (BECs) is a rich system to explore many interesting phenomena: collective excitations \cite{maddaloni2000collective}, pattern formation \cite{hoefer2011dark}, phase separation \cite{papp2008tunable}, nonlinear excitations \cite{eto2016bouncing}, and the Kibble-Zurek mechanism \cite{nicklas2015observation}. Interspecies interactions, specially for isotopic condensate mixture, provide valuable insights into the fundamental aspects of quantum dynamics, some of which are experimentally tested, such as the mixtures of ${}^{85}Rb$-${}^{87}Rb$ \cite{papp2008tunable},  ${}^{160}Dy$-${}^{162}Dy$ \cite{tang2015bose}, ${}^{39}K$-${}^{41}K$ \cite{tanzi2018feshbach}, ${}^{87}Rb$-${}^{133}Cs$ \cite{mccarron2011dual}, ${}^{23}Na$-${}^{39}K$ \cite{schulze2018feshbach},  ${}^{41}K$-${}^{87}Rb$ \cite{burchianti2018dual}, and ${}^{23}Na$-${}^{133}Cs$ \cite{warner2021overlapping}.

Among the many fascinating phenomena observed, phase separation has been the subject of significant interest, where the miscible-immiscible transition is controlled by tuning the interspecies Feshbach resonance \cite{papp2008tunable,wen2020effects,kumar2017vortex}. One could note that the regime of separation for the ground state of the two species lies at the greater values of the interspecies interaction \cite{wen2012controlling,wen2020effects,cikojevic2018harmonically}. This is because, at higher values of interspecies interaction, the energy of the inhomogeneous state is lower than that of the homogeneous state, which favours the spatial separation of the components. The phase separation is also predicted for a BEC mixture under the Thomas-Fermi limit \cite{ao1998binary,riboli2002topology,jezek2002interaction}. Experimental observation of spatial separation is also achieved, but by neglecting the role of mass-imbalance between the isotopes \cite{papp2008tunable}. The miscible-immiscible transition is shown in the absence of an external trap, where the transition is governed by the strength of the interspecies interaction in comparison to the intraspecies interaction \cite{ao1998binary}.

On the other hand, the phase boundary of such transition is also shown to alter by changing the trap frequency \cite{navarro2009phase,merhasin2005transition,sabbatini2011phase}. The external trap, being the most favourable physical quantity to control the dynamics of a BEC, drives a quick emergence of various technological applications \cite{bloch2012quantum,ryu2015integrated,gajdacz2014atomtronics}. A large amount of literature exists towards the theoretical and experimental studies of efficient trap engineering in BEC \cite{lesanovsky2007time,henderson2009experimental,nath2014bose,raghav2022tunneling,nath2020exact,halder2021exact,basu2022nonlinear}. A ring-shaped waveguide is one of the most useful traps in $2$D, which is formed by overlaying a blue-detuned laser in the middle of harmonic confinement \cite{ryu2007observation,naik2005optically}, where the radius of the ring can be efficiently controlled. BEC inside such a ring waveguide manifests several interesting physics. Such a ring trap is a fundamental constituent of atomtronics, which is aimed at manipulating ultracold matter for quantum technological applications \cite{RevModPhys.94.041001}. The phenomenon of fractional revivals (FR) is recently reported in this system \cite{bera2020matter} and is a very well-studied effect in diverse quantum systems in their time evolutions \cite{averbukh1989fractional,robinett2004quantum,parker1986coherence,banerji2006role,ghosh2006mesoscopic,ghosh2009sub,roy2009sub,ghosh2012coherent,ghosh2014enhanced}.

In this work, we investigate the mixing-demixing transitions of a binary mixture of trapped Bose-Einstein condensates driven by fractional revivals. The underlying model involves the dynamics of a binary BEC inside the ring-shaped waveguide with ${}^{85}Rb$ and ${}^{87}Rb$ isotopes as an example. The inclusion of interspecies interaction makes the FR-physics more rigorous and interesting due to the emergence of two time-scales, which are influenced by each other. We conduct a systematic study of the time evolution and identify experimentally feasible ranges of trap parameters and interspecies interaction strengths that lead to separation of the condensates. Additionally, we analyze how the mass ratio between the species affects the average separability, revealing a universal behavior across different interspecies interaction strengths.

The paper is organized as follows. The next section deals with the model of interacting BEC mixture in a ring trap, along with the numerical technique to be adopted. Section III includes a detailed analysis of the combined dynamics through a modified FR phenomenon, where the influence of interspecies interaction on the individual time scales becomes apparent. A range of interspecies interactions is proposed in Sec.IV, which later suggests physical situations for spatial separation through autocorrelation function. In Sec.V, we quantify the degree of separation at different time instances and identify the favourable situations for greater separability. A precise parameter domain is also revealed for ring radius and interspecies nonlinearity strength. Visualization of condensate densities at the identified instances clearly manifests the spatial separation of the two components, which validates our model. In Sec. VI, we analyze the behavior of average separability as a function of mass ratio for various interspecies interaction strengths. The paper concludes in Sec.VII with a summary and possible implications.

\section{Basic Formulation of the System and the Methodology}

Let us consider that ${}^{85}Rb$ and ${}^{87}Rb$ have atomic masses $m_i$ and, number of atoms $N_i$, for $i=1,2$, respectively. Both the components have their respective intra-species scattering lengths, $a_{11}$ and $a_{22}$, whereas inter-species coupling is governed by the scattering length $a_{12}$. This inter-species interaction can be tuned to desired values through the inter-species Feshbach resonance  \cite{burke1998prospects,pappFeshbach,dong2016observation,fukuhara2009all,tanzi2018feshbach}. The three-dimensional (3D) mean-field Gross Pitaevskii Equation (GPE) describes the dynamics of a two-component BEC in a ring trap. To make the equation dimensionless, we scale the position, time, and energy by $l_{0}$, $1/\omega_{0}$ and ${\hbar \omega_{0}}$, respectively. Here, $l_{0}=\sqrt{\hbar/2m_{1}\omega_{0}}$ is the harmonic oscillator length of the ODT in the radial direction. $\omega_{0}$ is taken as $\omega_{1,r}$, the trap frequency of ODT as experienced by ${}^{85}Rb$ in radial direction.

Preparation of the initial isotope mixture is crucial. The two components are initially confined in a magnetic trap, where the ${}^{85}Rb$ atoms are in the $|{F=2,m_{f}=-2}\rangle$ state and the ${}^{87}Rb$ atoms are in the $|{F=1,m_{f}=-1}\rangle$ state. This configuration relies on their magnetically sensitive states for initial confinement and cooling. Once the atoms reach sufficiently low temperatures, they are transferred to an optical dipole trap (ODT), where further cooling is performed to create a dual-species BEC \cite{papp2008tunable}. The optical dipole trap confines both species in a spin-independent manner, enabling fine control over their spatial overlap. The initial overlapping cloud of mixed isotope BEC is prepared with a width $d_0=4.505 \mu$m in the ODT of radial frequency $\omega_r=2\pi\times2.9$ Hz. Achieving a perfectly overlapped initial state can be challenging due to the requirement of fine tuning of interspecies Feshbach resonance. However, by implementing fast magnetic sweeps through the resonance point faster than the time scale of molecule formation and atom loss due to inelastic collisions \cite{pappFeshbach}, one could potentially improve stability in interspecies overlap.

The 3D wavefunction of the system is obtained by reducing the dimensionless 3D-GPE to the quasi-2D GPE by factorizing as follows $\psi_i(x,y,t)\phi_i(z)$. Here $\phi_i(z)=(\frac{\lambda_i}{\pi})^{\frac{1}{4}}e^{-\frac{1}{2}\lambda_iz^2}$ is the wavefunction along z-direction for $i^{th}$ component and $\lambda_i=\frac{\omega_{i,z}}{\omega_{i,r}}\gg 1$ is the aspect ratio for the given transverse and longitudinal frequency. The mass dependence in the wavefunctions $\phi_1(z)$ and $\phi_2(z)$ can be absorbed by choosing suitable aspect ratios, $\lambda_i$. To maintain a quasi-2D configuration, the condition $\lambda_i>>1$ is satisfied with $\lambda_1=44.83$ and $\lambda=44.31$. These values ensure that the transverse wavefunctions remain mass independent \cite{kumar2020mass}. All the physical quantities are chosen as follows: $N_{1}=N_{2}=10^{4}$; $m_{1}=85\;$a.u., $m_{2}=87\;$a.u., $l_{0}=d_{0}=4.505 \mu$m;  $a_{11}=a_{22}=2.698\times10^{-9}\;$m \cite{papp2008tunable}. The quasi-2D GPE is written as follows \cite{pethick2008bose,adhikari2001coupled,salasnich2014localized}.
\begin{equation}
    i\frac{\partial \psi_{i}}{\partial t}=[\mathcal{L}_{i}+\mathcal{N}_{i}]\psi_{i},
    \label{GPE}
\end{equation}
where $\psi_{i}\equiv\psi_{i}(x,y,t)$ and $\mathcal{L}_{i}=-{\frac{m_{1}}{m_{i}}\nabla_{x,y}}^{2}$ with $\nabla_{x,y}^2=\frac{\partial^2}{\partial x^2}+\frac{\partial^2}{\partial y^2}$. The second term,
$\mathcal{N}_{i}=\sum_{j=1} ^{2}g_{ij}|\psi_{j}|^2+V_i(x,y)$, contains the tunable quantities, such as the nonlinearities,
$g_{ij}=\frac{\sqrt{2\pi\lambda_i}m_{1}(m_{i}+m_{j})a_{ij}N_{j}}{m_{i}m_{j}l_{0}}$ and the ring trap ($V_i(x,y)$) for for $i=1,2$. In this case, both the species experience the same potential in the form of two Gaussian beams \cite{moscatelli2007far}:
\begin{eqnarray}
	V_i(x,y)=V_{G}\Bigg[1-e^{-\frac{2(x^2+y^2)}{\alpha^2}}\Bigg]+V_{0}e^{-\frac{2(x^2+y^2)}{\sigma^2}}. \label{Ring}
\end{eqnarray}	
Here $\alpha$, $\sigma$ and $V_G$, $V_{0}$ are the waists and amplitudes of the two Gaussian beams. For $\alpha>>\sqrt{2(x^2+y^2)}$, the potential reduces to
\begin{eqnarray}	
    V_i(x,y)=\frac{1}{4}\rho_{i}\omega_i^2(x^2+y^2)+V_{0}e^{-\frac{2(x^2+y^2)}{\sigma^2}}.
\end{eqnarray}
Here, $\omega_{i}=\sqrt{\frac{8V_{G}}{\rho_i \alpha^2}}$ is the frequency of the harmonic trap and $\rho_{i}=\frac{m_i}{m_1}$, such that $\rho_1\omega_1^2=\rho_2\omega_2^2$. So the ring is approximated as a combination of a 2D harmonic potential and a Gaussian barrier. The oscillations along the radial direction are suppressed by placing the condensate in the exact minimum ($r_{0}$) of the potential \cite{zhang2017spin}.\\
\textit{Numerical Method}:
For our simulation, we have taken experimentally feasible values for the amplitudes and waists of the two Gaussian beams, $V_G=1512.5\hbar\omega_0$, $V_0=223\hbar\omega_0$, $\alpha=110l_0$ and $\sigma=10l_0$ \cite{kohnen2008ultracold,baumert2013dipole,yin2006realization,davidson1995long}. We adopt the Split Step Fourier Method (SSFM) \cite{weideman1986split, taha2005parallel,wang2007time} for numerically solving the system, where both the parts of the dynamical equation (Eq.\ref{GPE}) are treated separately. The first term is evolved in the momentum space, and the second term, involving the nonlinearity and the trap, is evolved in the coordinate space \cite{weideman1986split}. The $x$- and $y$-coordinates are equally divided into $512$ grids with a step size of $0.1841$. The step size for time is $0.0915$ with a total of $16384$ grids up to the second revival time.

\begin{figure*}[ht]
    \centering
     \includegraphics[width=14 cm]{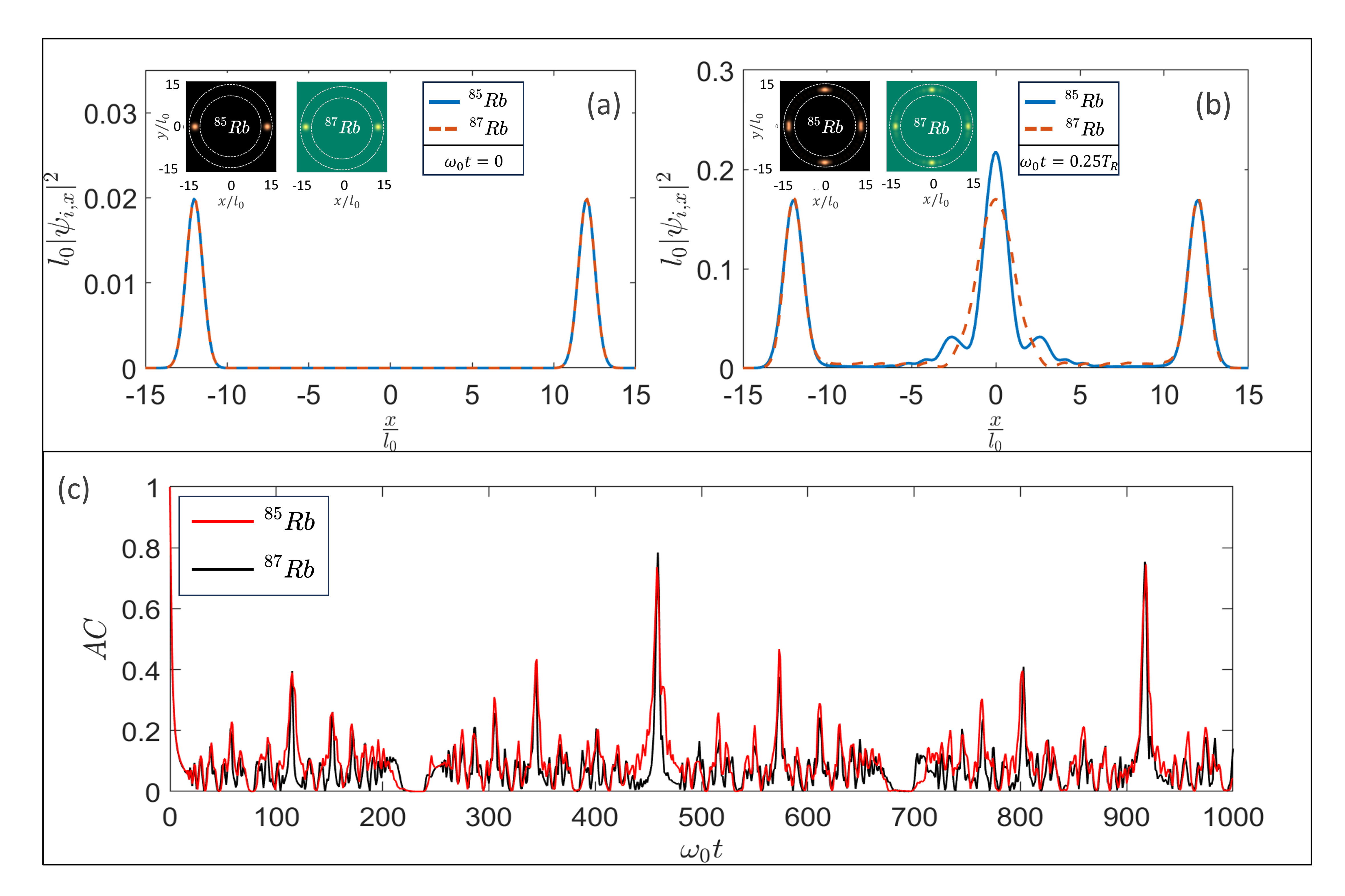}
    \caption{The $x$-axis projected density of the BEC mixture of ${}^{85}Rb$ and ${}^{87}Rb$ with interspecies interaction strength $a_{12}=1.0\;a_{11}$, (a) $t=0$ and (b) the quarter revival time, $t=T_{R}/4$. The 2D densities of the two isotopes are shown separately as insets in (a) and (b). The autocorrelation functions of the two interacting species are shown in (c). The ring radius is $r_{0}=12l_{0}$. $x$ and $y$ are scaled by $l_{0}=4.505\;\mu$m and, $t$ is scaled by $1/\omega_{0}=54.88\;$ms.}
    \label{mixed}
\end{figure*}

\section{Dynamics of the Condensate Mixture}

We consider that the condensates of the two isotopes initially coexist in the form of a binary peak with waist, $d_{0}=1\;l_{0}$, and at the diametrically opposite points of the ring with coordinates, $(\pm r_{0},0)$. The mixed cloud disperses along the ring waveguide in clockwise and anti-clockwise directions. They will start interfering at $(0,\pm r_{0})$, and also continue to spread further.

It is important to note that a single component BEC (not the system under study) in a ring trap, after some specific time interval, revives in its initial position and shape. The time when the condensate replicates the initial configuration is termed the revival time, $T_{R}$. This revival phenomenon has been thoroughly examined recently, where the exact revival time is given by $T_{R}=\pi r_{0}^2$ in the units of $\frac{1}{\omega_0}$ \cite{bera2020matter}. Moreover, at some specific fractions of this revival time ($\omega_0t=T_{R}p/q$), several mini replicas of the initial condensate are formed, and this phenomenon is known as fractional revivals (FR), where $p$ and $q$ are mutually prime integers and decide the number of splits \cite{bera2020matter}. According to the model, at time $\frac{T_{R}}{4}$ ($p=1$, $q=4$), a single initial cloud will split into two, and a dual initial cloud will split into four daughter condensates. Moreover, two components of a binary BEC with no interspecies interaction will independently show FR and follow the above model, which is not the case for a binary mixture of BECs where interspecies interactions are considered.

To display the resultant cloud of the binary mixture in the presence of interspecies interactions $a_{12}=1.0\;a_{11}$, we show the $x$-axis projected density of the initial cloud at Fig. \ref{mixed}(a) and the cloud at quarter revival time $T_{R}/4$ in Fig.\ref{mixed}(b). The $x$-axis projected density is obtained by integrating out the $y$ coordinate. It is apparent that the condensates of both the isotopes are mixed and their densities overlap to a greater extend. This becomes further clarified from the $2$D condensate densities of ${}^{85}Rb$ and ${}^{87}Rb$ shown in the insets of Fig.\ref{mixed}(a) and \ref{mixed}(b). These densities look identical and not distinct, resulting in a miscible cloud. To separate two constituent clouds of the interacting isotopes, we need to unwind the physics of miscibility and then try to devise a way. The well-known autocorrelation (AC) function helps us in the first stage. AC is the modulus square of the inner product of the initial and the temporally evolved wavefunctions, having mathematical definition as
\begin{equation}
    |A_i(t)|^{2}= {|\int_{-\infty}^{\infty}\int_{-\infty}^{\infty}\psi_i^{*}(x,y,0)\psi_i(x,y,t)dx dy|^2}.
\end{equation}
For a dispersing cloud, the AC function decays with time to manifest, gradually decreasing fidelity with the initial structure. However, the existence of significantly dominant peaks have different physics and are also observed in Fig.\ref{mixed}(c) for the mixed cloud with interspecies interaction  $a_{12}=1.0\;a_{11}$. Periodic AC peaks with nearly one magnitude are the signature of revivals, whereas there are other periodic peaks of lower magnitudes, which are known as FR instances. Overall, the AC time-series provides us with comparable characteristic time scales for both the components in the presence of interspecies interactions, due to which the components remain indistinguishable.

Therefore, Fig.\ref{mixed}(a) to Fig.\ref{mixed}(c) suggest designing a physical situation when AC peaks will get separated in time during some FR instances to distinguish and measure two components from their mixture. The key physical parameters to control the dynamics are the radius of the ring trap ($r_{0}$), interspecies interaction ($a_{12}$) and time. It is clear from Fig.\ref{mixed} that the interspecies interaction, $a_{12}=1.0\;a_{11}$, and ring radius, $r_{0}=12 l_{0}$, do not help in spatial separation during the whole temporal dynamics. Below, we will explore the influence of interspecies interaction on the individual revival dynamics of the isotopes.

\subsection{Revival Dynamics of the Mixture without and with Interspecies Interactions}

For a two-component BEC, we have two-time scales, one for each component. First, we will discuss about the expression for revival times of the two species at zero interspecies interaction. It is known that an initial Gaussian wave packet of two components with width $w_{in,1}=w_{in,2}=w_{in}$, centred at $(r_{0},0)$, propagates along the ring and interferes with itself at $(-r_{0},0)$. Let us consider $s$ as the arclength coordinate with periodic boundary conditions, which is equivalent to the circumference of the ring. The initial cloud, placed at the edges of $s$, is taken as $\phi^{a}_{i}(s,t)$ and $\phi^{b}_{i}(s,t)$, since the edges follow periodic boundary conditions. Here, $a$ and $b$ denote the two edges of the $s$ coordinate. In order to obtain the interference term, we consider the total density due to the interference of these wavepackets:
\begin{eqnarray}
	n_i(s,t)=|\phi^{a}_{i}(s,t)+\phi^{b}_{i}(s,t)|^2,\nonumber\\
	n_i(s,t)=|\phi^{a}_{i}(s,t)|^2+|\phi^{b}_{i}(s,t)|^2+2Re[\phi^{a}_{i}(s,t){\phi^{b}}^{*}_{i}(s,t)].
\end{eqnarray}
In the case of zero interspecies interaction, we can treat the two components independent and the interference term of each component is given by
\begin{eqnarray}	
	I_{i}= 2Re[\phi^{a}_{i}(s,t){\phi^{b}}^{*}_{i}(s,t)],\nonumber\\
    2Re[\phi^{a}_{i}(s,t){\phi^{b}}^{*}_{i}(s,t)]\propto \cos{\Bigg(\frac{2Dts}{\rho_{i}w_{in}^2w_{t,i}^2}\Bigg)}.
    \label{interferenc}
\end{eqnarray}
The initial clouds $\phi^{a}_{i}(s,t)$ and ${\phi^{b}}_{i}(s,t)$, having equal width $w_{in}$, are separated by a distance $D$. $w_{t,i}$ is their widths at a later time $t$. These widths are related by \cite{pethick2008bose}
\begin{eqnarray}
w_{t,i}=\sqrt{w_{in}^2+\Bigg(\frac{2t}{\rho_{i}w_{in}}\Bigg)^2}.
\end{eqnarray}
The interference maxima are obtained by taking the argument of the cosine in Eq.\ref{interferenc} as $2\pi n$.  Hence, the effective fringe separation of the interference pattern is obtained from the distance between two consecutive interference maxima:
\begin{equation}
    \Delta s_{i}= \frac{4\pi t}{\rho_{i}D}
\end{equation}
At revival, the fringe separation becomes the integral multiples of $2 \pi r_{0}$ ($2 \pi r_{0} \times p$) due to the circular geometry of the ring, where $p$ denotes the winding number. Since the initial separation $D$ between the wave packets is $2 \pi r_{0}$, the revival time for ${}^{85}Rb$ and ${}^{87}Rb$ in the absence of interspecies interaction are calculated to obtain
\begin{eqnarray}
    T_{R,i}=\pi \rho_{i}r_{0}^2 \times p \label{times}.
\end{eqnarray}
The difference of the revival times of ${}^{85}Rb$ and ${}^{87}Rb$ in absence of interspecies interaction becomes
\begin{equation}
    \Delta T_{R}=\Bigg(\frac{m_{2}}{m_{1}}-1\Bigg) \pi r_{0}^2 .\label{difference}
\end{equation}

\begin{figure}[ht]
    \centering
    \includegraphics[width=8.5 cm]{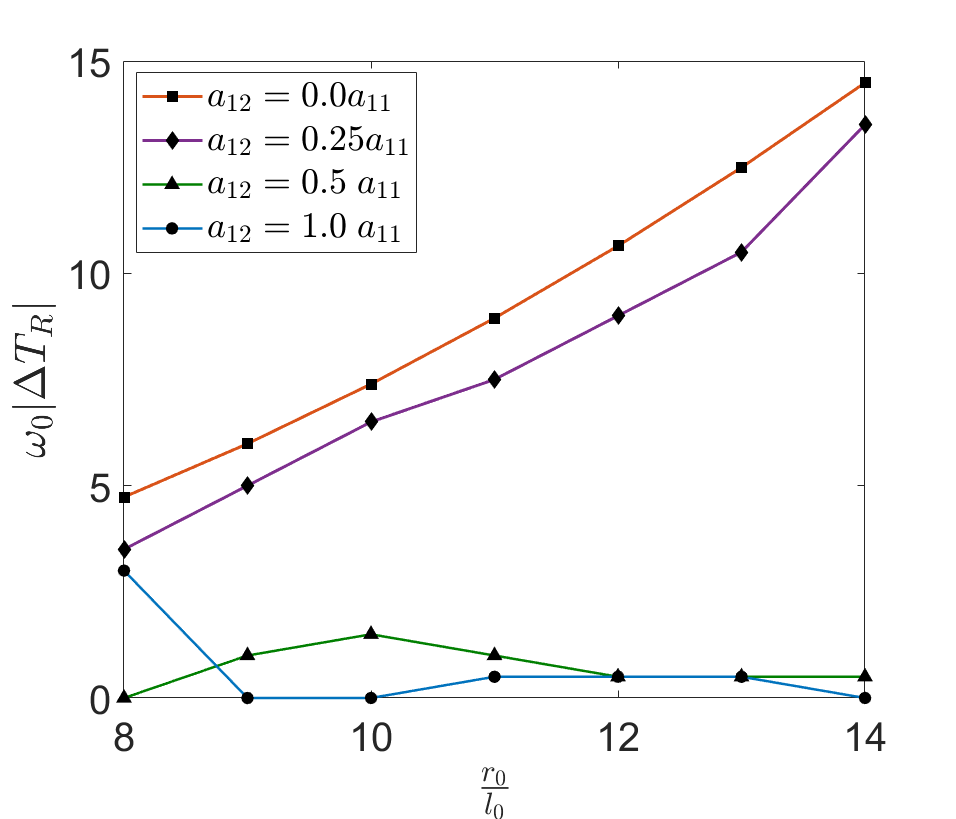}
    \caption{Difference in the revival times of the two components $\Delta T_{R}$, for various radii are shown. The solid line with square is in the absence of interspecies interaction, whereas the solid lines with diamond, triangle and circle are in presence of interspecies interaction, $a_{12}=0.25a_{11}$, $a_{12}=0.5a_{11}$ and $a_{12}=1.0a_{11}$, respectively. Here, radius $r_{0}$, revival time $T_{R}$, and interspecies interaction $a_{12}$ are scaled by $l_{0}=4.505\;\mu$m, $1/\omega_{0}=54.88\;$ms, and ${a_{11}=2.698\;}$nm, respectively.}
    \label{time_r}
\end{figure}

It is clear from the above equation that the difference in the revival time scales for two noninteracting components is merely due to the mass-imbalance of the two species. The solid line with squares in Fig. \ref{time_r} shows $\omega_{0}|\Delta T_{R}|$ with the radius of the ring in the absence of interspecies interaction. This variation is quite straightforward from the above analytical expression. However, the presence of interspecies interaction makes the variation quite nontrivial, and it doesn't follow Eq.\ref{difference}. In this case, $\omega_{0}|\Delta T_{R}|$ is plotted after numerically solving the dynamical equation and by finding the revival times for an interacting mixture, as depicted by the solid line with diamond, triangle and circle in Fig. \ref{time_r} for nonzero interactions, $a_{12}=0.25a_{11}$, $a_{12}=0.5a_{11}$ and $a_{12}=1.0a_{11}$, respectively. When $a_{12}=0$, the difference in revival times for ${}^{85}Rb$ and ${}^{87}Rb$ isotopes scales quadratically with $r_0$. As $a_{12}$ slightly increases, this quadratic relation starts to break and at around $a_{12}=0.5 a_{11}$, the revival times of the two species becomes close to each other. This explains why we didn't have spatial separation for $a_{12}=1.0\;a_{11}$ in Fig.\ref{mixed}. The important points from Fig. \ref{time_r} are i) the nonuniform time scale variation with the ring radius and ii) the possibility of influencing the time scale variation with respect to the different interspecies interaction strengths.

\section{Identifying Appropriate Interspecies Interaction for Spatial Separation}

\begin{figure*}[ht]
    \centering
    \includegraphics[width=14 cm]{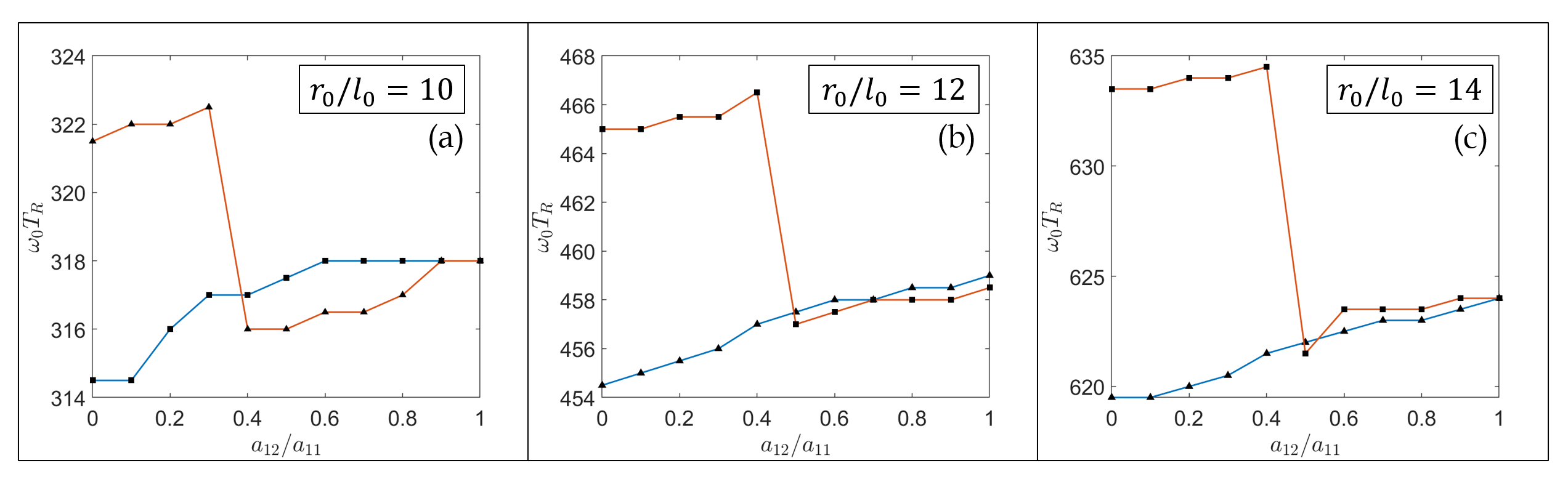}
    \caption{Variation of revival time for ${}^{85}Rb$ (line with squares) and ${}^{87}Rb$ (line with triangles) with interspecies interaction $a_{12}$, for fixed ring radii, (a) $r_{0}=10 l_{0}$ (b) $r_{0}=12 l_{0}$ and (c) $r_{0}=14 l_{0}$. Here, $r_{0}$, $T_{R}$, and $a_{12}$ are scaled by $l_{0}=4.505\;\mu$m, $1/\omega_{0}=54.88\;$ms, and ${a_{11}=2.698\;}$nm, respectively.}
    \label{time_int}
\end{figure*}

We have seen that the binary mixture with interspecies interaction doesn't follow Eq.[\ref{times}] and both the components have their revival times altered due to interaction strength for a given external trap. We evaluate the revival times of the $Rb$ isotopic BEC for a wide range of interaction strengths with constant ring radii and depict them in Fig.(\ref{time_int}). A merging of the time scales is observed at higher interspecies interactions. This behaviour is also seen for other fractional and revival times of the two components at greater interspecies interaction. We have shown the merging of timescales for ring radii $r_{0}=10l_{0}$, $r_{0}=12l_{0}$ and, $r_{0}=14l_{0}$, in the Fig.\ref{time_int}(a), \ref{time_int}(b) and \ref{time_int}(c), respectively.

The interesting point to gain here is the wide difference in the time scales of the interacting components for a significant range of interspecies interaction. Moreover, the greater the radius, the greater the difference in the revival times at $a_{12}=0$ as shown in Fig.\ref{time_r}. This difference in the time scales is definitely the situation where the mixed condensate cloud should manifest a separation.  The spatial separation is not possible for interspecies interaction strength $a_{12}\gtrsim 0.5 a_{11}$, due to insignificant differences in the revival times of the two species. This range of interspecies interactions that allow unmerged revival time scales increases with the decrease of the inverse mass ratio, $\rho_{2}^{-1}$. A smaller $\frac{m_1}{m_2}$ exhibits a greater range of interspecies interactions that facilitates the separation of the two components of the BEC. Specifically, for Rubidium isotopes, with $\rho_{2}^{-1}=0.9770$, the interspecies interaction spans from $0$ to $0.5a_{11}$.\\
Therefore, for spatial separation, one needs to choose $1)$ an interspecies interaction strength for which the revival times of the two components are fairly separated and $2)$ a ring radius for which the difference in the revival times, in the absence of the interspecies interaction strength, is considerably large.
\begin{figure*}[ht]
    \centering
    \includegraphics[width=14 cm]{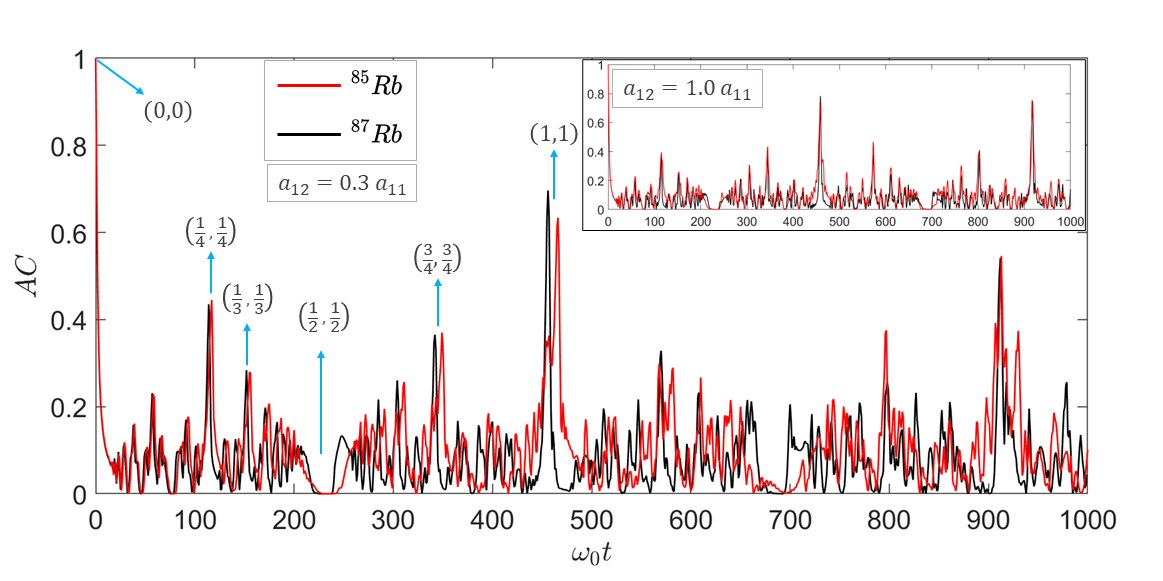}
    \caption{Autocorrelation function of two initial condensates of a coupled BEC with interspecies interaction $a_{12}=0.3\;a_{11}$. Fractional revival instances are indicated by pair notations. The inset shows the autocorrelation function of the two species at $a_{12}=1.0\;a_{11}$ for comparison. The ring radius is taken as $r_{0}=12l_{0}$ and $t$ is scaled by $1/\omega_{0}=54.88\;$ms.}
    \label{AC}
\end{figure*}

To examine whether such a situation is indeed favourable for separating the BEC of two miscible components, we choose a set of preferred parameters: $r_{0}=12 l_{0}$ and interspecies interaction strength, $a_{12}=0.3a_{11}$. We repeat the AC function plot with these parameters in Fig. \ref{AC} for ${}^{85}Rb$ (solid line) and ${}^{87}Rb$ (dotted line) interacting condensates, where every peak corresponds to a fractional revival time, and we compare it with the previous AC function (Fig.\ref{mixed}(d)) plot, given in the inset of Fig. \ref{AC}. AC peaks are seen to separate, unlike in the previous case. The FR times of both the species are indicated by the notation, $(t_1,t_2)$, such that $(\frac{3}{4},\frac{3}{4})$ corresponds to $\frac{3}{4}^{th}$ revivals of both, ${}^{85}Rb$ and ${}^{87}Rb$. The spacing of the peaks increases with time, thereby making it possible to identify time instances where the difference in the AC functions of the two species is maximum. As discussed earlier, the greater the difference in the two AC functions, the lesser the spatial overlap between two components.
Two condensate wavefunctions will become distinguished or separately measured when the pair of peaks are separated from each other. In other words, the maximum of one AC peak should coincide with the minimum of the other AC peak. This, being a dynamical system, such typical situation has to be maintained for separation. The pair of peaks $(1,1)$ is one such instance. Hence, our methodology is indeed helpful in choosing the optimal values of $r_{0}$ and $a_{12}$ for which the isotope separation is possible.

\section{Separability of the two-components}

From the autocorrelation functions of the two species, one could notice the separation of the components. However, it requires careful analysis of how much they are separated and at what times for efficient implementation of the isotope separation scheme. To obtain the degree of separation, we define a quantity called 'Separability ($S$)' between the two components in the mixed BEC:
\begin{figure*}[ht]
    \centering
    \includegraphics[width=14 cm]{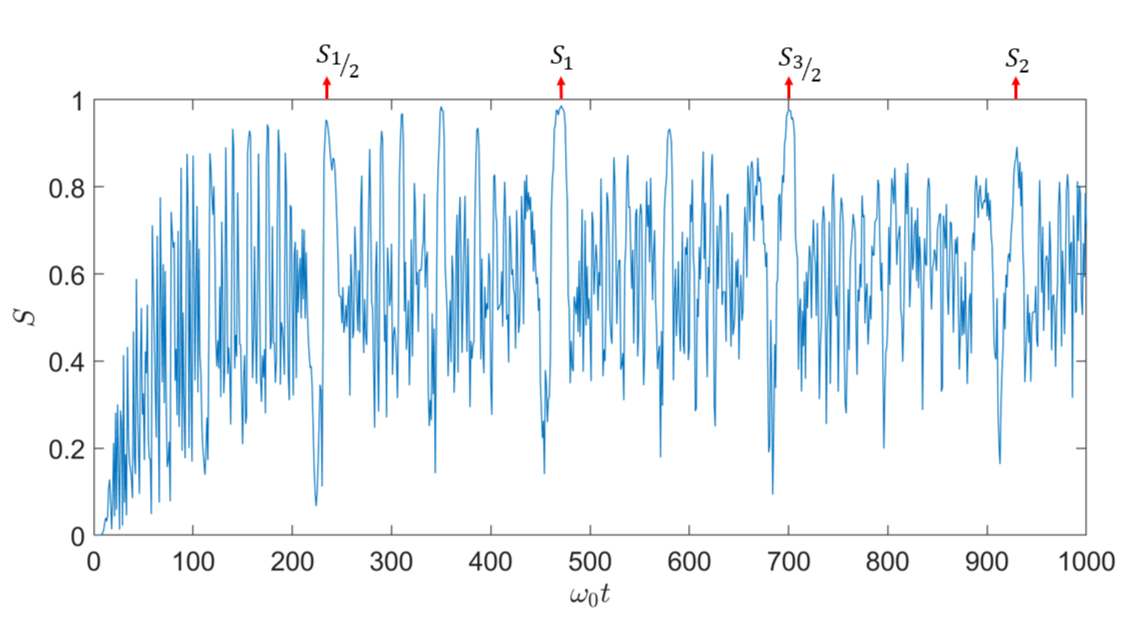}
    \caption{Variation of the Separability with Time. The prominent separability peaks are indicated, such as $S_{\frac{1}{2}}$ denotes the time near the $1/2$-th fractional revival of ${}^{87}Rb$. The interspecies interaction is taken as $a_{12}=0.3 a_{11}$ and the ring radius is $r_{0}=12l_{0}$. $t$ is scaled by $1/\omega_{0}=54.88\;$ms.}
    \label{Separability}
\end{figure*}
\begin{eqnarray}
    &S=1-\Delta,\\
    &\Delta=\frac{[\int_{-\infty}^{\infty}\int_{-\infty}^{\infty}|\psi_{1}(x,y)|^2|\psi_{2}(x,y)|^2 dx dy]^{2}}{\int_{-\infty}^{\infty}|\psi_{1}(x,y)|^4 dx dy \int_{-\infty}^{\infty}|\psi_{2}(x,y)|^4 dx dy}. \label{delta}
\end{eqnarray}
The term, $\Delta$, is given by the square of the inner product of the probability densities of ${}^{85}Rb$ and ${}^{87}Rb$, weighted by the product of $4^{th}$ order moment of the inner product \cite{chen2019immiscible,jain2011quantum,suthar2017characteristic,halder2023two}. The numerator helps in amplifying the tiny variations in the overlap of the wavefunctions of two species. The separability takes the value from $0$ to $1$, where $0$ implies zero separation and $1$ corresponds to a $100\%$ separation. We choose the parameters of Fig. \ref{AC} for calculating $S$ and presented in Fig. \ref{Separability}. A separability peak close to $1$ will be the desired situation. For times up to $1000$ in the unit of $\frac{1}{\omega_{0}}$, we could choose four prominent  peaks, $S_{\frac{1}{2}}$, $S_{1}$, $S_{\frac{3}{2}}$ and $S_{2}$, occurring closer to $\frac{1}{2}$, $1$, $\frac{3}{2}$ and $2$ of the revival time of ${}^{87}Rb$.

The times corresponding to $S_{\frac{1}{2}}$, $S_{1}$, $S_{\frac{3}{2}}$ and $S_{2}$ are $12.84s,\;$$25.85s,\;$$38.42s$ and, $51.04s$, respectively. At $S_{\frac{1}{2}}$, $S_{1}$ and $S_{\frac{3}{2}}$, the condensate densities of the two species have $>95 \%$ separation and at $S_{2}$, the two species have $89\%$ separation. The exact values of the percentage of separation of the two components of $Rb$ isotopic mixture for the above Separability peaks are given in Table.\ref{table1}.

\begin{table}[ht]
\centering
    \begin{tabular}{ | c | c | c | c | c | }
    \hline
        \boldmath \hspace{.4cm} Peak \hspace{.4cm} & \hspace{.4cm} $S_{\frac{1}{2}}$ \hspace{.4cm} & \hspace{.4cm} $S_{1}$ \hspace{.4cm} & \hspace{.4cm} $S_{\frac{3}{2}}$ \hspace{.35cm} & \hspace{.4cm} $S_{2}$ \hspace{.4cm}\\ \hline
        Time (s) & 12.84 & 25.85 & 38.42 & 51.04 \\  \hline
        Separability & 95.42$\%$ & 98.65$\%$ & 97.62$\%$ & 89.19$\%$ \\ \hline
    \end{tabular}
    \caption{Percentage of separation of the two isotopes for the four Separability peaks and their corresponding times from Fig. \ref{Separability}. The interspecies interaction is taken as $a_{12}=0.3 a_{11}$ and the ring radius is $r_{0}=12l_{0}$.}
    \label{table1}
\end{table}

In addition, the quality of spatial separation of the two-component BEC will be relying on the following factors too: $1)$ at higher evolution time, one could observe an overall decay of the autocorrelation function (Fig.\ref{AC}) due to dispersion and hence, sooner is better for spatial separation; $2)$ a broader temporal width of the separability peak is preferable, as it will provide better time window in which the two components remain separated in the experiment.\\

\begin{figure}[ht]
    \centering
    \includegraphics[width=8.5 cm]{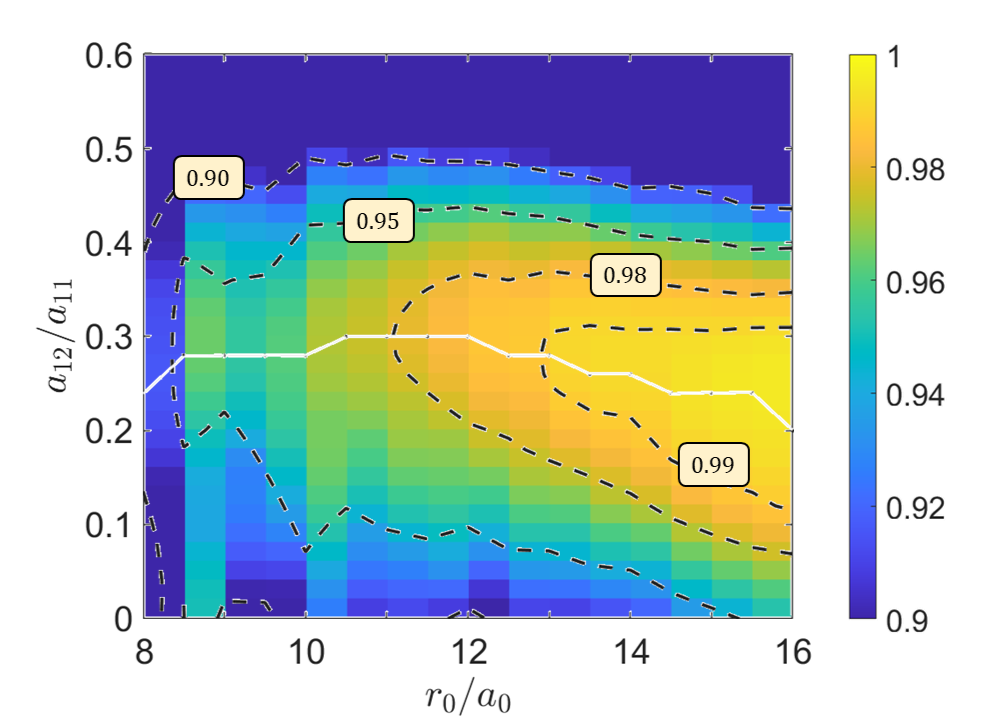}
    \caption{The preferred parameter regions for spatial separation, following the Separability, $S_{1}$, as an example. The interspecies interaction $a_{12}$ and ring radius $r_{0}$ can be chosen as per the need, where the white dotted line indicates the maximum separability. We also draw the regions for various percentages of separation, $90\%,\; 95\%,\; 98\%$ and $99\%$, by dashed contours. The ring radius is scaled by $l_{0}=4.505\;\mu$m, and the interspecies interaction is in the unit of $a_{11}=2.698\;$nm. The color bar indicates the value of $S_{1}$ separability.}
    \label{separability_parameters}
\end{figure}

\textit{\textbf{Parameter Contours for Maximal Spatial Separation}}: In our work, the separability of the $Rb$ isotopic BEC is tuned by two physical parameters, the radius of the ring $r_{0}$ and the interspecies interaction $a_{12}$. The separability peaks in Fig. \ref{Separability} offers the times for maximal spatial separations, designated by $S_{\frac{1}{2}}$, $S_{1}$ and $S_{\frac{3}{2}}$ and having $>95 \%$ separation. The exact values of the percentage of separation are given in Table.\ref{table1}. These, along with the above points ($1$ and $2$), suggest one of the most favourable instances as $S_{1}$ with a wider separability peak and $98.65\%$ separation. We identify the parameter contours, comprising of $r_{0}$ and $a_{12}$, for the instance, $S_{1}$, and depict it in Fig. \ref{separability_parameters}.

The diagram highlights the regions above $90{\%}$ separability. Though the whole highlighted region provides a wide parameter range for spatial separation, one can further improve it by choosing greater seperability, as shown by different contours. The white dotted line in the middle indicates the maximum separability value for both parameters.The magnetic field $B$ is related to values of interspecies interaction strength $a_{12}$ through the Feshbach Resonance is given by \cite{Li2008}:
\begin{equation}
    a_{12}=a_{bg}\Bigg(1-\frac{\Delta B}{B_{0}-B}\Bigg) \label{FReqn}
\end{equation}

For the mixtures of ${}^{85}Rb-{}^{87}Rb$, the background scattering length is $a_{bg}=11.27\times 10^{-9}$m, the Feshbach Resonance peak position is $B_{0}=265.42\;G$ and the width is taken $\Delta B=5.8\;G$ as per the experiment \cite{pappFeshbach,Li2008}.
It is interesting to note that the maximum separability line lies within the window $0.2a_{11}>a_{12}>0.3a_{11}$ of interspecies interaction strength. We also draw the regions for various percentage yields, $90\%,\; 95\%,\; 98\%,$ and $99\%$, by dashed contours. For a desired separation line, the favourable values for the radius and interspecies interaction can be chosen in the experiments. \\

\begin{figure*}[ht]
    \centering
    \includegraphics[width=14 cm]{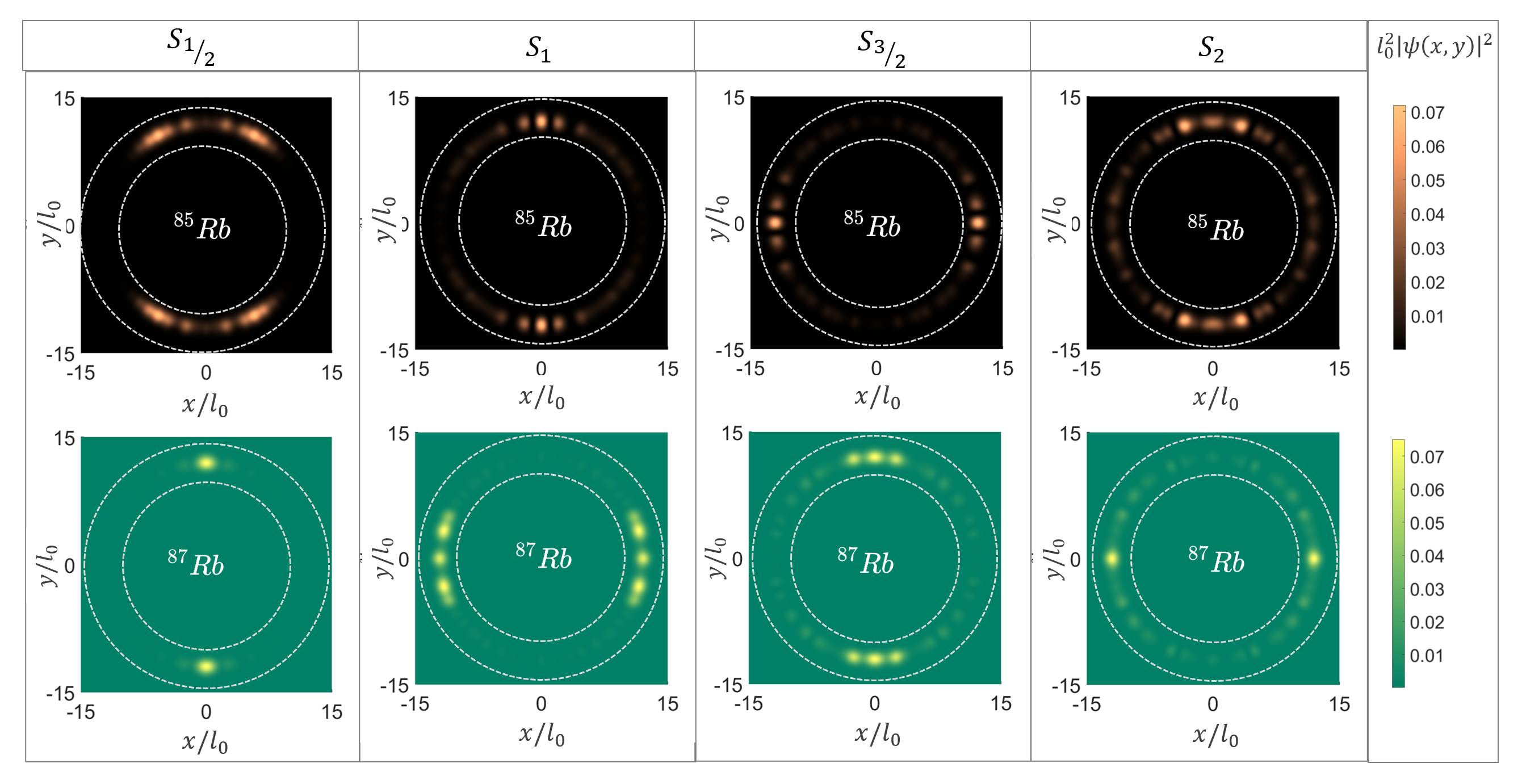}
    \caption{Condensate densities of both the isotopes, ${}^{85}Rb$ and ${}^{87}Rb$, separately plotted at times, $12.84\;$s,\;$25.85\;$s,\;$38.42\;$s, and $51.04\;$s, corresponding to $S_{\frac{1}{2}}$, $S_{1}$, $S_{\frac{3}{2}}$ and $S_{2}$, respectively. Spatial coordinates are scaled by $l_{0}=4.505\;\mu$m. The interspecies interaction is taken as $a_{12}=0.3 a_{11}$, the ring radius is $r_{0}=12l_{0}$.}
    \label{density_sep}
\end{figure*}

\textit{\textbf{Condensate Densities Upon Spatial Separation}}: One can visualize the individual condensate of the binary BEC for any set of parameters, described in the parameter contour plot in Fig. \ref{separability_parameters} to verify their spatial separation. Here, we will choose one of the values, such as the interspecies interaction is $a_{12}=0.3 a_{11}$ and the ring trap radius is taken as $r_{0}=12l_{0}$ in Fig. \ref{density_sep}. Condensate densities of both the isotopes, ${}^{85}Rb$ and ${}^{87}Rb$, separately plotted at times, $12.84\;$s,\;$25.85\;$s,\;$38.42\;$s, and $51.04\;$s, corresponding to $S_{\frac{1}{2}}$, $S_{1}$, $S_{\frac{3}{2}}$ and $S_{2}$, respectively. These instances, near multiples of half revival, do not further split the initially considered dual clouds and, hence, show nice separation in the preferred parameter window. At $S_{\frac{1}{2}}$, the isotopes are separated with respect to their common centre of mass and remain separated for a shorter time interval, as reflected from the narrow width of the separability peak for $S_{\frac{1}{2}}$ in Fig. \ref{Separability}. It is fascinating to observe that each of the other pairs of plots for ${}^{85}Rb$ and ${}^{87}Rb$ condensates in Fig. \ref{density_sep} manifests orthogonal positioning with respect to each other, implying a clear separation inside the ring trap.

These observations emphasize the dependence of separability dynamics on both the interspecies interaction strength and the system’s physical parameters. To generalize these results, it is worth analyzing the behavior of average separability as a function of mass ratio for various interspecies interaction strengths for binary BECs.

\section{Universal Miscibility Behaviour of Binary BECs for Different Interspecies Interactions}

\begin{figure*}[ht]
	\centering
	\includegraphics[width=14 cm]{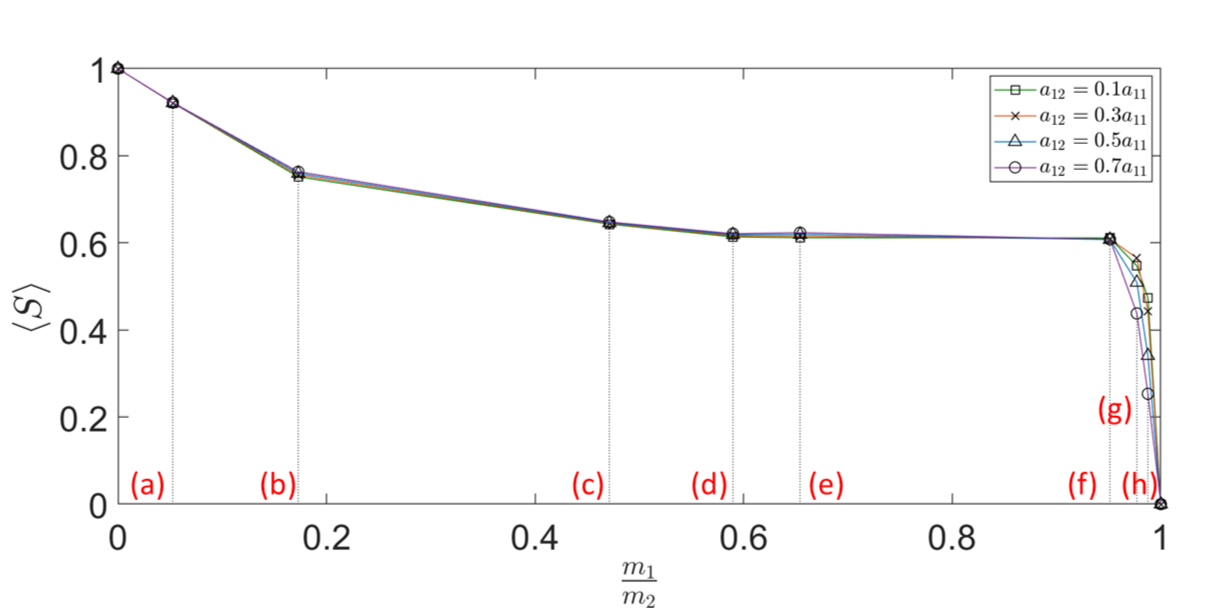}
	\caption{Average Separability, taken over the time window $t=0$ to $t=500 \omega_0^{-1}$, vs mass ratio for varying interspecies interaction strengths. The mass ratios used here are (a) ${}^{7}$Li-${}^{133}$Cs, (b) ${}^{23}$Na-${}^{133}$Cs, (c) ${}^{41}$K-${}^{87}$Rb, (d) ${}^{23}$Na-${}^{39}$K, (e) ${}^{87}$Rb-${}^{133}$Cs, (f) ${}^{39}$K-${}^{41}$K, (g) ${}^{85}$Rb-${}^{87}$Rb and (h) ${}^{160}$Dy-${}^{162}$Dy. The interspecies interaction strength is in the unit of $a_{11}$, the ring radius is $r_{0}=12l_{0}$, where  $l_{0}=4.505\;\mu$m.}
	\label{S_avg}
\end{figure*}

The average Separability $\left\langle S \right\rangle$, evaluated over the time window $t=0$ to $t=500 \omega_0^{-1}$, as a function of the mass ratio $\frac{m_1}{m_2}$, exhibits a universal behavior under the varying interspecies interaction strengths. An average separability value of
$\left\langle S \right\rangle=1$ indicates that the two components remain completely separated throughout the evolution,  while $\left\langle S \right\rangle=0$ corresponds to complete overlap. An intermediate value, such as 0.5, signifies that the components remain, on average, $50\%$ spatially separated. Figure \ref{S_avg} shows the dependence of $\left\langle S \right\rangle$ on the mass ratio for different interspecies interaction strengths, $0.1a_{11}$, $0.3a_{11}$, $0.5a_{11}$, and $0.7a_{11}$, represented by lines with square, cross, triangle, and circle markers, respectively. For small mass ratios (regions (a) and (b)), the separability remains close to $1$, reflecting robust spatial demixing between the components. As the mass ratio increases, $\left\langle S \right\rangle$ decreases gradually, indicating a transition toward increasing miscibility.

It is interesting to observe a saturated behaviour of the average separability for mass ratios, $0.58-0.95$ (regions (c), (d) and (e)), highlighting a balance between spatial mixing and separation. Beyond this region, the average separability experiences a sharp decay towards the zero average separability. This is attributed to the nearly equal revival time scales of the two components (Eq.\ref{difference}). Hence, it is more interesting to see how both the components can be separated in this region and thus, we have chosen the isotopic BEC-mixture of ${}^{85}$Rb-${}^{87}$Rb for demonstration, which has the average separability close to $55\%$. Since isotopic mixtures have mass ratios close to $1$, their average separability does not exceed $60\%$.

\section{Conclusion}
We investigated the role of interspecies interaction strengths and trap parameters on the miscibility dynamics of mass-imbalanced binary BEC. The difference in the fractional revival time scales of the two interacting components is used to identify the specific time instances of spatial separation. We have quantified the degree of spatial separation of the two components by a metric called 'Separability'. The plot of separability \textit{vs} time gives us specific time-instances of maximal spatial separation. The times corresponding to $S_{\frac{1}{2}}$, $S_{1}$ and $S_{\frac{3}{2}}$ are identified, where the condensate densities of the two species have $>95 \%$ separation and $89\%$ separation at $S_{2}$, for $r_{0}=12l_{0}$ and $a_{12}=0.3a_{11}$. The percentage of separation is addressed up to $\sim99\%$. The physically supported parameter contours offered a wide range of trap and interspecies nonlinearity values under the present scheme. The domain of interspecies interaction strength is unique ($a_{12}<a_{11}$) in the context of spatial separation in binary BECs, as compared to the past works \cite{wen2012controlling,wen2020effects,cikojevic2018harmonically}. Lastly, we demonstrated the universal behaviour of mass-imbalanced binary BECs under varying interspecies interaction strengths. Our findings show that low mass ratios exhibit robust spatial demixing, while mass ratios close to $1$ result in identical revival dynamics, leading to increased miscibility between the components. We have demonstrated results in this region for isotopic mixture of ${}^{85}Rb$ and ${}^{87}Rb$ condensates. 

\section*{Acknowledgements}
UR acknowledges the Science and Engineering Research Board, India for the support from the grant (project no. CRG/2022/007467).



\bibliography{REF}


\end{document}